\documentclass[sn-mathphys]{sn-jnl}

\jyear{2024}%

\begin{document}

\title[Article Title]{Inflation of 2D boundary ghosts and digital watermarking}

\author[1]{\fnm{Imants} \sur{Svalbe}}
\email{imants.svalbe@monash.edu}

\author[2]{\fnm{Rob} \sur{Tijdeman}}
\email{tijdeman@ziggo.nl}

\affil[1]{\orgdiv{School of Physics and Astronomy}, \orgname{Monash University}, \orgaddress{\city{Clayton}, \state{Victoria}, \country{Australia}}}

\affil[2]{\orgdiv{Mathematical Institute}, \orgname{Leiden University}, \orgaddress{\city{Leiden},\country{The Netherlands}}}

\abstract{Projection ghosts are discrete arrays of signed values positioned so that their discrete projections vanish for some chosen set of $n$ projection angles. Minimal ghosts are designed to be compact, with no internal pixels having value zero. Here we control the shape, number of boundary pixels and area that each minimal ghost encloses. Binary minimal ghosts and their boundaries can themselves be ‘inflated’ by tiling copies of themselves to make ghosts with larger sizes and different shapes, whilst still retaining the same set of $n$ zero projection angles. The intricate perimeters of minimal ghosts are formed by three strings of connected pixels that are defined (recursively) by the minimal projection angles. We show that large changes to the ghost areas can be made whilst keeping the length of their segmented perimeters fixed. These inflated boundary ghosts may prove useful as secure watermarks to embed into digital image data. For example, adding large but ‘thin’ (\unboldmath{$\pm1$}) boundary ghosts as watermarks in 8-bit integer image data is, in general, visually imperceptible. Ghost watermarks distribute marginal changes across all discretely projected views of the data, except for the $n$ ghost projection angles. Boundary ghosts may also help guide the selection of angles used to reconstruct images where the object domain is confined to oval shaped arcs.}



\maketitle

\section{Introduction}

Discrete projection of digital data and the reconstruction of images from sparse sets of projected data continues to find applications in low-dose imaging, file storage and data security. Projection ghosts play a large part in these applications \cite{guedonMojette}. Ghosts are digital images of signed values with the distinguishing property that their discrete projections are all zero for a pre-determined set of discrete angles \cite{Brunetti}.

The size and shape of a projection ghost image are important: both parameters determine if any image of a given size can be reconstructed exactly from a particular set of projections \cite{Katz}. If a ghost for $n$ projection directions fits everywhere inside the edges of any image, then that image cannot be reconstructed exactly using just those $n$ projections.

The minimum number of signed points necessary to compose a ghost in $n$ directions is equally important, being $2n$ only for $n = [1,2,3,4,6]$, see \cite{Gritzman_n12346}, \cite{Alpers_n12346}. The minimum number of ghost points jumps to $12$ for $n=5$, $20$ for $n=7$ and  grows rapidly beyond that. For \emph{periodic} discrete arrays, there are minimal binary ghosts made with just $2n$ points for any $n > 0$ projection directions in $p \times p$ arrays for prime $p > 2n$ \cite{svalbe_normand}. The periodic wrapping of projected rays provides the extra degrees of freedom not available for the \emph{aperiodic} projections of $n \times m$ data considered here.

This paper is about control of the size, shape and number of non-zero points in projection ghosts for $n$ directions, and adds new variations of minimal ghosts beyond those reported in \cite{CekoGhosts}. In that work, ghosts, images made with pixel values of $\pm1$, were termed ‘maximal’ when they contained $2^n$ simply connected pixels and had zero-sums for $n$ projection directions.  As simply connected shapes of $\pm1$ points also enclose the minimum number of pixels, we will instead here call them ‘minimal ghosts’. Concise explanatory material on projection ghosts, the discrete Radon transform and especially on the construction of 'maximal' or 'minimal' ghosts and their boundaries can be found in \cite{CekoThesis}.

The size of an $n$-direction projection ghost must grow larger as $n$ increases. The symbol $U_n$ is used here for compact minimal ghosts with zero-sum projections in $n$ directions. For ghosts built on a regular lattice, if there are no gaps between any of the $\pm1$ ghost elements, then the perimeter of a 2D minimal area ghost generates a minimal boundary ghost (and the surface of a $3D$ minimal volume projection ghost is a minimal $3D$ boundary ghost \cite{Ceko3D}). The symbol $V_n$ is used here for $2D$ minimal boundary ghosts in $n$ directions. Each minimal boundary ghost satisfies the following strict conditions:

1.	has a connected boundary of pixels, with adjacent values that alternate between $\pm1$, that encloses an empty interior.

2.	all discrete projections of the boundary points, for $n$ selected directions, are everywhere exactly zero.

We show in this work that, for any chosen number $n$ of 2D ghost projection directions, minimal boundary ghosts can themselves be tiled in the 2D plane to form ghosts of almost any larger size and shape, whilst retaining the same zero-sum projection properties. Such tilings are here called  ‘inflated boundary ghosts’ and are denoted by symbol $W_n$.

As a particular example, we show that compact tilings of a minimal boundary ghost (packed tightly around the original ghost), can enclose areas from four to seven times larger where, surprisingly, all these ghosts have the same constant perimeter.

 The perimeters of all boundary ghosts $V_n$ are comprised of three parallel pairs of curved segments. Each perimeter segment can be found at displacements that are linear combinations of the projection directions that define the minimal ghost. Inflated boundary ghosts can themselves be tiled at multiple levels of inflation along any of the six possible symmetry-axis directions. This flexibility permits construction of a large variety of connected boundary ghosts of different sizes and shapes, all having the same $n$ zero projections.

Inflated ghosts, with different sizes and aspect ratios, may be useful to embed into digital data as an imperceptible watermark \cite{SchyndelOsborneWaterMark}. More efficient watermarking schemes may be now needed to counter the widely available artificial intelligence programs that create 'fake' videos and documents. A projection-based watermark is different to, for example, the existing spectral or correlation-based schemes \cite{SchyndelWaterMark}. Adding a ghost $W_n$ as a watermark into data marginally changes the content of all projected views of the data, except at the known zero-projection angles of the added ghost.

Some specialised forms of projective imaging, such as potential or acoustic tomography, restrict the shapes of the imaged objects to be confined to lines along the arc of a circle \cite{RTforArcData}, or to parts of an annular circle or sphere \cite{PlanarcircularRT}. The zero-sum boundary ghosts may in these cases help to determine if certain projections cannot contribute to an exact reconstruction, as their curved domain may support one or more forms of boundary ghost.

Section \ref{sec:binary_ghosts} introduces the construction of binary ghosts from an initial tile of simply connected pixels with values of $\pm1$. Section \ref{sec:boundary_ghosts} constructs examples of ghosts that skirt the perimeter of the binary ghost area, enclosing no internal signed pixels. Section \ref{sec:defineUn} reviews the method from \cite{CekoGhosts} to construct binary minimal ghosts (here called $U_n^a)$, as well as new related ghosts (here called $U_n^{a'}$, $U_n^b$). Section \ref{sec:defineVn} shows example minimal boundary ghosts, $V_n^a$, $V_n^{a'}$ and $V_n^b$, $V_n^{b'}$ for $n= 8$. Section \ref{sec:inflated ghosts} describes inflation of boundary ghosts by tiling pairs or multiple copies of ghosts $U_n$ or $V_n$. A brief analysis is given of the properties of $V_n$ and $W_n$ for different values of $n$. Section \ref{sec:watermarking} shows a proposed example of digital image watermarking that uses inflated boundary ghosts. Section \ref{sec:flexibility} discusses options that vary the shape and size of connected boundary ghosts in $n$ directions, with a view to making their shapes unique and secure for use as signatures in watermarking.

\section{Binary ghosts}\label{sec:binary_ghosts}

In this paper, a ghost is a map $g: \mathbb{Z}^2 \to \mathbb{Z}$ such that all the line sums in certain directions vanish. We identify the points of $\mathbb{Z}^2$ with the distinct 1 by 1 squares surrounding them and call them pixels. In images we use white to denote pixels with value $+1$, black for pixel value $-1$ and grey for pixel value $0$.

A $2D$ $n$-directional ghost is constructed as a planar image, starting from an arbitrary finite set $S$ in $\mathbb{Z}^2$ where each pixel of $S$ has an integer value assigned. All other pixels of $\mathbb{Z}^2$ start with value $0$. Next a direction $v_1 =(p_1, q_1)$ is chosen, where $p_1$ and $q_1$ are co-prime integers. $S + v_1$ is a copy of $S$, shifted by $(p_1,q_1)$, where each pixel has the opposite value of the corresponding pixel in $S$. We add $S$ to its copy: let $S_1 = S \union (S+(p_1,q_1))$. Then the resulting discrete line sums taken along the direction $(p_1,q_1)$ are $0$.
Subsequently a direction $v_2 = (p_2,q_2)$ is chosen, where $(p_2, q_2)$ are co-prime integers, with $v_2$ not in the direction of $\pm v_1$. Negated values of $S_1$ are added to the values of the corresponding pixels of $S_1 + (p_2,q_2)$. Now  all the discrete line sums for both directions $v_1$ and $v_2$ are $0$.
The process of negation, displacement and addition is repeated for shift directions $v_i$, for $i = 3$ to $n$. Then all the line sums in the directions $v_1, v_2, \dots, v_n$ are $0$ and the result is a ghost $S_n$ for these $n$ directions.

The size of the image of any ghost $S_n$ is $a \times b$, where

\begin{equation}\label{eq:ghost_boxsize}
        a = 1 + \sum_{i=1}^n \mid{p_i}\mid,~~~ 
        b = 1 + \sum_{i=1}^n \mid q_i\mid, 
\end{equation}

\noindent{that is, one more than the sum of the absolute coordinate values of all the ghost directions $v_i$.}
\vskip.1cm
\noindent {\bf Example 2.1} Let $S = \{(0,0)\}$ and $v_1, v_2, \dots, v_{10}$ be given by 

$$[0,-1; 1,0; 1,-1; 1,1; 1,-2; 2,-1; 1,-4; 4,-1; 2,-3; 3,-2].$$

\begin{figure}
    \centering
    \includegraphics[width=0.33\linewidth]{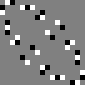}
    \caption{Image of binary ghost, $S_{10}$. The 20 pixels with value $+1$ and the 20 pixels with value $-1$ together make projections that are everywhere zero for the 10 directions of Example 2.1.} 
    \label{fig:n10ghost40points}
\end{figure}

\noindent Figure \ref{fig:n10ghost40points} shows an example of a binary $(\pm 1)$ ghost, $S_{10}$, comprised of 40 non-zero pixels in a box of size $17$ by $17$ pixels. This image has zero discrete line sums for the 10 directions given in Example 2.1. \qed

The ghost $S_{10}$ is comprised of the smallest possible number of pixels, However it is not a minimal \emph{area} ghost, nor a \emph{perimeter} or boundary ghost, as it has internal pixels with value zero and $\pm1$.

Suppose each pixel of the initial tile $S$ has value $+1$ or $-1$ and, after each addition, no overwriting of any ghost values occurs. Then, after $n > 0$ ghost directions have been added, the ghost will contain $\mid \emph{S} \mid \times 2^n$ pixels, half of them with pixel value $1$, the other half with value $-1$. Such a ghost is called a maximal binary ghost, following the method used in \cite{CekoGhosts}. Here $ \mid \emph{S} \mid$ denotes the cardinality of $S$. The next step is to construct ghosts with no over-writing \emph{and} no internal zero-valued pixels.

A sequence of points $a_0, a_1, \dots, a_n$ in $\mathbb{Z}^2$ is called a 4-path if $\mid a_i - a_{i-1}\mid = 1$ for $i = 1, 2, \dots, n$ and an 8-path if $ 1 \leq \mid a_i - a_{i-1}\mid \leq \sqrt{2}$ for $i=1, 2, \dots, n$. 

Consider a finite set $S$ of distinct points in $\mathbb{Z}^2$. The set $S$ is called 4-connected if it is possible to go from any point of $S$ to any other point of $S$ via a 4-path in $S$. Similarly $S$ is called 8-connected if it is possible to go from any point of $S$ to any other point of $S$ via an 8-path in $S$. A set $S$ is called simply connected if both $S$ and its complement $\mathbb{Z}^2 \setminus S$ are 4-connected. It means that the union of the pixels of $S$ has no holes (here meaning pixels with value zero).

Here we measure the area of a binary ghost by counting the number of pixels its perimeter encloses. A maximal binary ghost, having $\mid \emph{S} \mid \times 2^n$ pixels \emph{and} a simply connected domain, occupies a fixed count of pixels in the plane, even for different minimal ghost shapes that share the same $n$. We here call these ghosts 'minimal', with a view to the following sections that their 'area' grows by multiples of $2^n$ pixels when these minimal ghosts are tiled.

\begin{figure}
    \centering
    \includegraphics[width=0.6\linewidth]{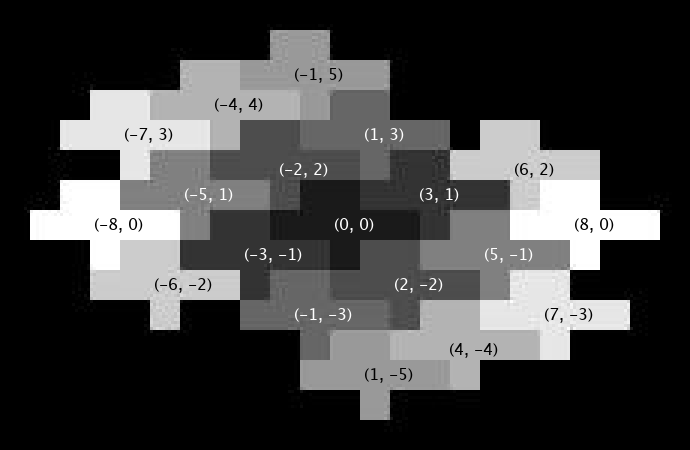}
    \caption{Tiling pattern of the array T from Table \ref{tab:tileT_5x3example}, centred at location $(0, 0)$, shifted by vectors ($i, j$).}
    \label{fig:tilingbyT}
\end{figure}

By a (linear) tiling we mean a set $T \subset \mathbb{Z}^2$ and vectors $t_1, t_2 \in \mathbb{Z}^2$ such that every point in $\mathbb{Z}^2$ can be written in a unique way as 
$$ t + t_1u_1 + t_2 u_2,$$ with $t \in T$ and $t_1, t_2 \in \mathbb{Z}$. $T$ is assumed to be simply connected and is called a tile.
Figure \ref{fig:tilingbyT} shows a tiling of the plane by the shape $T$, as defined in Example 2.2 below, for the labelled vector shifts $(i,j).$

We choose a pair $u_1, u_2 \in \mathbb{Z}^2$ such that both $T + u_1$ and $T + u_2$ are tiles adjacent to $T$. Every discrete rectangle with sides parallel to the axes is an example of a tile.
\vskip.1cm
\noindent {\bf Example 2.2} Consider the tile $T$ of $8$ pixels in a box of size $5 \times 3$, labelled by the values $1$ as shown in Table \ref{tab:tileT_5x3example};
\vskip.1cm
\begin{table}[h]
    \centering
    \begin{tabular}{cccccc}
          & 1 & 1 &  &  \\
         1 & 1 & 1 & 1 & 1 \\
          &  & 1 &  &  \\
        \end{tabular}
    \caption{A tile $T$ generating a linear tiling with $t_1 = (3,1)$ and $t_2 = (2,-2)$. Fig. \ref{fig:tilingbyT} presents an image of the tile $T$ from Table \ref{tab:tileT_5x3example} as the central tile (shaded in black) with the corresponding copies of $T$ tiled at shifts $(p_i, q_i)$ as labelled.}
    \label{tab:tileT_5x3example}
\end{table}

To form a tiling (with no gaps) the directions $u_1, u_2$ of the adjacent tiles can be chosen as $\pm (3,1)$ and $\pm (2,-2)$. Necessarily the determinant of these directions is $\pm 8$, where 8 is the number of pixels of the tile. We can assign the pixels within $T$ to have arbitrary values $1$ or $-1$. For example, we choose the pattern of signs shown in Table \ref{tab:signed_tileT_5x3}.

\begin{table}[h]
    \centering
    \begin{tabular}{cccccc}
          & 1 & -1 &  &  \\
         1 & 1 & -1 & 1 & -1 \\
          &  &  1 &  &  \\
        \end{tabular}
    \caption{The tile T from Table \ref{tab:tileT_5x3example} with signed pixels.}
   \label{tab:signed_tileT_5x3}
\end{table}

We take the union of $T$ and one of its adjacent tiles, $T + (3,1)$ as the new tile and write $v_1 = (3,1)$. In the added tile, we assign its pixel values to be the opposite of those in the corresponding pixels of the original tile $T$. Their union is a new simply connected tile of $16$ non-zero pixels in a box of size $8 \times 4$ with values $1$ and $-1$.

\begin{table}[h]
    \centering
    \begin{tabular}{cccccccc}
           &  &  &  & -1 & 1 &  &  \\
           & {\bf 1} & {\bf -1} & -1 & -1 & 1 & -1 & 1 \\
         \bf 1 & \bf 1 & \bf -1 & \bf 1 & \bf -1 & -1 &  &  \\
           &  & {\bf 1} &  &  &  &  &  \\
        \end{tabular}
    \caption{Signed tile $T$ from Table \ref{tab:signed_tileT_5x3}, shown in bold, with a negated copy of $T$ shifted by $(3,1)$. The line sums in the direction $(3,1)$ are all $0$.}
    \label{:tile3x5plus_shiftedT}
\end{table}

\noindent Note that, by construction, the line sums in direction $(3,1)$ are all $0$, as shown in Table \ref{:tile3x5plus_shiftedT}. The adjacent tiles are given by shift vectors 
$$(6,2),(1,3),(-2,2),(-3,-1),(2,-2),(5,-1).$$ 

We can continue this procedure by adding, for example, the tile shifted by $v_2 = (2,-2)$, to form a new tile of $32$ pixels. The new tile consists of the tiles indicated in Fig. \ref{fig:tilingbyT} by $(0,0)$, $(3,1)$, $(2,-2)$, $(5,-1)$. Here all the line sums in the directions $(3,1)$ and $(2,-2)$ are zero.

Proceeding in this way we get, after the $n^{th}$ extension, a minimal binary ghost comprised of $\mid T \mid \times ~2^n$ pixels, half of them with value $1$ and half of them with value $-1$. Here $\mid T \mid = 8.$
In the $n$ chosen directions all the line sums are zero. \qed

\section{Boundary ghosts}\label{sec:boundary_ghosts}

A boundary ghost arises from a defined starting tile $T$. A shift direction $(a,b)$ is specified, with the property that in $T$ all the pixel locations that differ by a multiple of $(a,b)$ have the same value. We give an example.
\vskip.1cm
\noindent{\bf Example 3.1} 
As starting tile $T$, we choose the $3\times 6$ array of non-zero pixels, as shown in Table \ref{tab:tile_T_signed_3x6}.
 
\begin{figure}
    \centering
    \begin{minipage}{0.5\linewidth}
      \centering
      \scriptsize{(a)}\\  
       \includegraphics[width=0.6\linewidth]{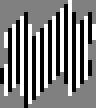}
    \end{minipage}%
   \begin{minipage}{0.5\linewidth} 
   \centering
   \scriptsize{(b)}\\
   \includegraphics[width=0.6\linewidth]{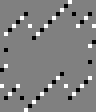}
   \end{minipage}
   
    \caption{(a) Tiling of the plane by copies of the $12$ pixel starting tile $T$ from Table \ref{tab:tile_T_signed_3x6}, with alternating signs, shifted by $v_n$. The array has zero projections in $5$ directions. The image size is $24 \times 27$. (b) Boundary ghost derived from (a) by applying boundary vector $(0,1)$. The image size is $24 \times 28$. The boundary ghost (b) has zero projections in $6$ directions and consists of $52~ (= 2 \times 24 + 4)$ pixels, covering an area of $410~ (=12 \times 2^5 + 52:2)$ pixels.}
    \label{fig:n5_a)minimal_b)boundary_ghosts}
    
\end{figure}

\begin{table}[h]
    \centering
        \begin{tabular}{ccc}
          &   & 1 \\
          & -1 & 1 \\
         1 & -1 & 1 \\
         1 & -1 & 1 \\
         1 & -1 &  \\
         1 &   &  \\
         \end{tabular}
    \caption{A starting tile $T$ used to construct a binary minimal ghost. $T$ has size $3\times6$ with pixel values $\pm1$, for which $(a,b) = (0,1)$.}
    \label{tab:tile_T_signed_3x6}
\end{table}
\noindent          
We define a sequence of shift vectors, $v_n$, for $\{v_n\}_{n=1}^{\infty}$ in $\mathbb{Z}^2$, from starting vectors $v_1 = (1,-3),~v_2 = (-3,-3)$ and recursion

\begin{equation} \label{eq:Vn1}
v_n = v_{n-1} - 2v_{n-2}~~(n=3, 4, \dots).
\end{equation}
\noindent 
Thus $v_3= (-5,3)$, $v_4 = (1,9)$, $v_5 = (11, 3)$. After each shift the ghost doubles in size. After five steps we have the simply connected ghost shown in Fig. \ref{fig:n5_a)minimal_b)boundary_ghosts}(a).

Note that in each column of the ghost there are either no $+1$'s or no $-1$'s (and, being simply connected, $T$ has no internal zeros). Adding another shift in the vertical direction, $v_6 = (0,1)$, results in each column of the tiling now having internal pixel values of $0$'s, with a single $+1$ and $-1$ at the ends of each column. In this way we obtain a binary ghost for $6$ directions: $[1,-3; -3,-3; -5,3; 1,9; 11,3; 0,1]$. This ghost has the property that it only has non-zero values if it is a vertical boundary point of the minimum area ghost. We call such a binary ghost a boundary ghost. The boundary ghost of Fig. \ref{fig:n5_a)minimal_b)boundary_ghosts}(a) is shown in Fig. \ref{fig:n5_a)minimal_b)boundary_ghosts}(b). \qed 

In Example 3.1 the boundary direction used was $(a, b) = (0,1)$. In the rest of the paper, boundary directions $(0,1)$ (for uniformly signed vertical columns) and $(-1,1), (1,1)$ (for uniformly signed diagonals), will be used and compared.

In order to have a boundary direction $(0,1)$ one has to secure that no column in the minimal ghost has pixels with $+1$ and $-1$ values. One way to achieve this is to start with a tile which has this property. Next, choose $v_2$ to have an odd first coordinate and then apply Eq.(\ref{eq:Vn1}) for $v_n$ with $n \geq 3$. These choices have the additional property that the successive tiles are all simply connected and keep the size of the boundary ghosts small.

Instead of Eq.\eqref{eq:Vn1}, we could apply the recursion of Eq. \eqref{eq:Vn2} that uses a positive summation: 

\begin{equation} \label{eq:Vn2}
v_n = v_{n-1} + 2v_{n-2}~~(n=3, 4, \dots).
\end{equation}

However, using Eq.\eqref{eq:Vn2} can lead to long and thin binary minimal ghosts shapes (with correspondingly near-parallel boundary ghost directions). For large $n$, the vectors $v_n$ from Eq.\eqref{eq:Vn2} fall closer and closer to either side of a fixed direction, for example, $(1,2)$.

\section{Minimal binary ghosts}\label{sec:defineUn}
A minimal binary ghost begins with the smallest possible starting tile, a single pixel, $T = {(0,0)}$. Then, for simply connected tiles, $\mid v_1\mid ~= 1$, $\mid v_2 \mid~ \leq \sqrt{2}$.
We next construct several possible alternative shapes for these minimal binary ghosts. The direction vectors $v_n$ for all $n > 2$ satisfy Eq.(\ref{eq:Vn1}). In all cases, we choose the value of the starting single pixel $(0,0)$ to be $+1$.

For minimal ghosts with alternating vertical \emph{columns} of pixels with uniform sign, the initial vector directions used in the recursion Eq.\eqref{eq:Vn1} are chosen as $v_1 = (1,0),~v_2 = (1,1)$ for $U_n^{a}$ and $v_1 = (1,0), ~v_2 = (-1,1)$ for $U_n^{a'}$. 

The minimal ghost $U_n^b$ has alternating \emph{diagonals} of pixels with uniform sign. Here the initial vector directions used in the recursion Eq.\eqref{eq:Vn1} are $v_1 = (1,0),~v_2 = (0,1)$. Changing the signs or the ordering of those starting vectors produces either reflected or transposed versions of $U_n^{b}$ or $U_n^{b'}$.

\begin{figure}
    \centering
    \includegraphics[width=0.4\linewidth]{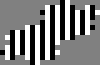}
    \includegraphics[width=0.4\linewidth]{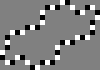}
    \caption{\hskip2cm(a) \hskip5cm(b)\\
    (a) Minimal ghost $U_7^a$ has image size $20 \times 13$ with a total of $2^7 = 128$ signed pixels, arranged in alternating columns of $\pm1$. (b) $V_8^a$, with boundary direction $(0,1)$, has image size $20 \times 14$ with $48$ signed pixels that define the perimeter of the shape in (a). The vertically-defined edges enclose a total area of $152$ pixels. The boundary ghost is an 8-connected curve where each adjacent pixel has the opposite sign.}
    \label{fig:ghosts_u71_v81}
\end{figure}

\begin{figure}
    \centering
    \includegraphics[width=0.4\linewidth]{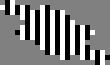}
    \includegraphics[width=0.4\linewidth]{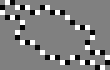}
    \caption{\hskip2cm(a) \hskip5cm(b)\\
    (a) Minimal ghost $U_7^{a'}$ has image size $22 \times 13$ with a total of $2^7 = 128$ signed pixels, arranged in alternating columns of $\pm1$. (b) $V_8^{a'}$, with boundary direction $(0,1)$, has image size $22 \times 14$ with $52$ signed pixels that define the perimeter of the shape in (a). The vertically-defined edges enclose a total area of $154$ pixels. The boundary ghost is an 8-connected curve where each adjacent pixel has the opposite sign.}
    \label{fig:ghosts_u71'_v81'}
\end{figure}

\begin{figure}
    \centering
     \includegraphics[width=0.32\linewidth]{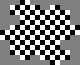}
    \includegraphics[width=0.32\linewidth]{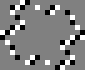}
    \includegraphics[width=0.32\linewidth]{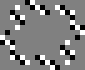}
    \caption{\hskip1cm(a) \hskip3cm(b) \hskip3cm(c)\\
    (a) Minimal ghost $U_7^b$ has image size $16 \times 13$ and consists of $2^7 = 128$ non-zero pixels with a chequered pattern of alternating $\pm1$ signs. (b) Ghost $V_8^b$, with boundary direction $(1,1)$, has image size $17 \times 14$ and consists of 48 pixels that enclose an area of 152 pixels. (c) Ghost $V_8^{b'}$, with boundary direction $(-1,1)$, has image size $17 \times 14$ and consists of 52 pixels that enclose an area of 154 pixels. Note that the adjacent pixels in all these ghost images have the opposite sign.}
    \label{fig:ghosts_u72_v82}
\end{figure}

\vskip.1cm
\noindent {\bf Example 4.1} Using the above starting directions and Eq.\eqref{eq:Vn1}, $v_n$ for $U_7^a$, $U_7^{a'}$ and $U_7^b$ are: 

$$U_7^a: [1,0;  1,1; -1,1; -3,-1; -1,-3; 5,-1;  7, 5],$$

$$U_7^{a'}: [1,0; -1,1; -3,1; -1,-1; 5,-3; 7,-1; -3, 5],$$

$$U_7^b: [1,0; 0,1; -2,1; -2,-1; 2,-3; 6,-1; 2, 5].$$
\vskip.1cm
Fig. \ref{fig:ghosts_u71_v81}(a) shows an image of the minimal ghost, $U_7^a$, with zero-sums for the first set of projection directions. Fig. \ref{fig:ghosts_u71'_v81'}(a) shows the minimal ghost $U_7^{a'}$. Fig. \ref{fig:ghosts_u72_v82}(a) shows the minimal ghost $U_7^b$.

$U_7^a, U_7^{a'}$ and $U_7^b$ ghosts are simply connected tiles, each is comprised of $2^7 = 128$ non-zero pixels. \qed

The angle set $U_n^a$ and $U_n^{a'}$ have shift vectors $v_i^a$ and $v_i^{a'}$ that are linear transformations of the corresponding shift vectors, $v_i^b$, for $U_n^b$. For vector $(p_{i}^b, q_{i}^b)$ in $U_n^b$, for $2<i\le n$

\begin{equation} \label{eq:pq1vspq2}
    (p_{i}^b, q_{i}^b) = (p_{i}^a - q_{i}^a, q_{i}^a) = (p_{i}^{a'} + q_{i}^{a'}, q_{i}^{a'}).
\end{equation}
\vskip.1cm

The structure of minimal binary ghosts $U_n^a$ that are generated by Eq.\eqref{eq:Vn1} for all $n$ has been treated in \cite{CekoGhosts}. In Section 5 of that paper, the ghost perimeter, here called $P_n$, has been analysed. It follows from Theorem 5 of \cite{CekoGhosts} that $P_n$ satisfies the ternary recursive relation 

\begin{equation} \label{eq:xnformula}
        P_{n} = P_{n-2} + 2P_{n-3} \ \ \ (n \geq 3). 
\end{equation}


\section{Minimal boundary ghosts} \label{sec:defineVn}

A minimal boundary ghost $V_n$ is derived from a minimal binary ghost $U_{n-1}$ (Fig. \ref{fig:ghosts_u71_v81}(a), Fig. \ref{fig:ghosts_u71'_v81'}(a) and Fig. \ref{fig:ghosts_u72_v82}(a) are example minimal ghost images). By adding the boundary direction as $n^{th}$ direction, all pixel values in the interior of $U_{n-1}$ become 0 and non-zero pixels only appear along the boundary of $U_{n-1}$.

The starting vectors for $U_n^{a'}$ are $(1,0), (-1,1)$ rather than $(1,0), (1,1)$ used for $U_n^{a}$. This changes the ghost vectors $v_n$ and hence alters the profile and perimeter for ghosts $V_n^{a'}$ compared to $V_n^{a}$. Minimal ghosts $U_n^{b}$ and $U_n^{b'}$ share the same set of start vectors, $(1,0),(0, 1)$, but have different boundary vectors, altering the profile and perimeter for ghosts $V_n^{b}$ compared to $V_n^{b'}$.

In \cite{CekoGhosts}, Theorem 7 proves that the minimal boundary ghost $V_n^a$ forms an 8-connected cycle of alternating pixel values $+1$ and $-1$. For $V_n^b$, the diagonal pixels are connected along the boundary by a string of alternating pixel values $+1$ and $-1$ at a distance at most $\sqrt{5}$. The number of boundary pixels, $B_n^a$, of $V_n^a$ satisfies the recurrence of Eq. (\ref{eq:xnformula}) with initial values $P_0^a=2, P_1^a=4, P_2^a =6$. This follows from \cite{CekoGhosts}, Theorem 8. The perimeter of boundary pixels for $V_n$ encloses an area $A_n$. That area is defined here as the number of pixels of the boundary ghost and all the pixels of $U_{n-1}$ enclosed by it in the boundary direction. The area, $A_n$ of $V_n$ is given by:

\begin{equation}\label{eq;vn_area}
    A_n = 2^{n-1} + P_n/2,    
\end{equation}

\noindent where $P_n$ denotes the number of pixels of the boundary ghost $V_n$. This follows from the fact that half of the boundary ghost pixels lie inside and half lie outside $U_{n-1}$.

The boundary direction required to turn $U_{n-1}^a$ into $V_{n}^{a}$ and to turn $U_{n-1}^{a'}$ into $V_{n}^{a'}$ is $(0, 1)$. From Eq.\eqref{eq:ghost_boxsize}, for the same $n$, we see that the boundary ghost $V_n^a$ fits into a smaller $P \times Q$ box than the box $P' \times Q'$ for $V_n^{a'}$ . Here $Q = Q'$, but $P'>P$. 

We use boundary direction $(-1, 1)$ for $V_n^b$ and $(+1, 1)$ for $V_n^{b'}$. Boundary ghost $V_n^b$ fits into the same sized $P \times Q$ box as does $V_n^{b'}$. 

 \vskip.1cm
\noindent{\bf Example 5.1} Fig. \ref{fig:ghosts_u71_v81}(b) shows the minimal boundary ghost $V_8^a$. Fig. \ref{fig:ghosts_u71'_v81'}(b) shows $V_8^{a'}$. Fig. \ref{fig:ghosts_u72_v82}(b) shows $V_8^b$,(c) shows $V_8^{b'}$.

We have $P_8^a = P_8^b = 48$, $A_8^a = A_8^b = 152$, whilst $P_8^{a'} = P_8^{b'} = 52$, $A_8^{a'} = A_8^{b'} = 154$. The boundaries of ghosts $V_8^a$ and $V_8^{a'}$  are 8-connected, whereas the ghosts $V_8^b$ and $V_8^{b'}$ have a small fraction of ghost pixels connected by vectors of length $\sqrt{5}$. \qed

\section{Inflating minimal boundary ghosts}\label{sec:inflated ghosts}

The area of any minimal ghost $U_{n-1}$ and the perimeter of the boundary ghost $V_n$ built from $U_{n-1}$ can be made larger by simply increasing $n$. However, some applications will require a variety of ghosts where $n$ should be kept relatively small and fixed. This section presents the major new findings of this paper. Here we examine many ways to change the size and shape of connected boundary ghosts $V_n$, for fixed $n$.

\subsection{Minimal ghost segments} \label{gendes}

As shown in Theorem 4 (cf. Fig. 4) of \cite{CekoGhosts}, every $U_{n-1}^a$ with $n>2$ can be surrounded by exactly six copies of itself. Then every $V_n^a$ has exactly six shifted copies with which it has a common segment. 

Each of these shifts merges a boundary ghost $V_n$ with one of its six surrounding tiles. Inflation is the process of merging a boundary ghost with one or more shifted copies so that the union is a boundary ghost for the same directions as are valid for the original boundary ghost. We denote the original boundary ghost by $W_n(1)$ and the resulting inflated boundary ghosts of $m$ tiles by $W_n(m)$.

There are two equivalent ways 'inflation' can be described. The first method assembles $m-1$ simply connected tiles of $U_{n-1}$ with itself. The sign of each added tile $U_{n-1}$ is selected to match the pattern of $\pm1$ pixels in the original tile. Next, the boundary vector is applied to reveal the perimeter of the expanded tiling as ghost $W_{n}(m)$. The second method assembles $m-1$ tiles of a boundary ghost $V_{n}$ where each shift makes opposite perimeter segments overlap and hence cancel (as opposite perimeter pixels always have opposite signs) to produce $W_{n}(m)$.

There are two ways to extend $U_{n-1}$ to a linear tiling. One tiling is to apply one or more of the six shift vectors of Eq. \eqref{eq:adjac} that are derived from Eq. \eqref{eq:Vn1},

\begin{equation} \label{eq:adjac}
  \pm w_1 = \pm 2v_n, ~\pm w_2 = \pm (v_n - 2v_{n-1}), ~\pm w_3 = \pm(v_n +2v_{n-1}).
\end{equation} 
\vskip.1cm

\noindent Note that $w_2 = v_{n+1}, ~w_3 = -v_{n+2}.$ The second, different tiling method is to use one or more of the six shift vectors of Eq. \eqref{eq:Vinflatedbyd},

\begin{equation}\label{eq:Vinflatedbyd}
    \centering
      \pm  w'_1 = \pm2v_{n}, ~\pm w'_2 = \pm2v_{n-1}, ~\pm w'_3 = \pm4v_{n-2}.
\end{equation}
\vskip.1cm

 For $W_n(1) = V_n$, the merging process can be repeated using any or all of the six shift directions: then $W_n(7)$ is a single larger ghost boundary formed by $V_n$ surrounded by all six copies of itself. An 'inflated' boundary ghost, $W_n(m)$, can be further inflated by overlapping shifts of itself onto its own opposite segments.
Adding an adjacent tile of a boundary ghost to itself annihilates the common segment(s) of their boundary, making the perimeter of the compounded ghost equal to twice the perimeter of the original ghost minus twice the number of pixels of the common (cancelled) segment(s).

\begin{figure}
    \centering
    \includegraphics[width=1.0\linewidth]{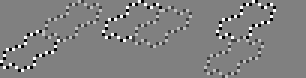}
    \caption{\hskip1cm(a) \hskip3cm(b) \hskip4cm(c)\\Images of inflated boundary ghost pairs $W_8^a(2)$. An image of the boundary ghost $V_8^a$ from Fig. \ref{fig:ghosts_u71_v81}(b), shown in bold, added pairwise to a copy of itself shifted by the vectors (a) $(14, 10)$, (b) $(10, -2)$ and (c) $(-4, -12)$. The ‘line’ or curved segment of internal boundary points for each shape is erased when the original and shifted boundary ghosts overlap with opposite pixel values. The erased segments consist of 6, 14, 4 pixels, respectively. The perimeters count 84, 68, 88, respectively, pixels, twice the perimeter of $V_8^a$ minus twice the length of the overlapped segments. The areas enclosed by each ghost pair are $298, 290, 300$ pixels respectively, being twice the area of $V_8^a$ (152 pixels) less the length of the overlapped segment.}
    \label{fig:V8asegments123}
    \end{figure}

\vskip.1cm
\noindent{\bf Example 6.1.} Three paired tilings of boundary ghosts $W_8^a(1)=V_8^a$ using Eq. \eqref{eq:Vinflatedbyd} are shown in Fig. \ref{fig:V8asegments123}(a), (b) and (c). Boundary ghost 
$V_8^a$ has perimeter of $48$ pixels and encloses aarea $152$ pixels. The common boundary segments $S_8^i$ are 6 pixels in (a), 14 pixels in (b) and 4 pixels in (c). These pixels disappear in the union. The perimeter of the new boundary ghosts $W_n^a(2)$ is therefore $2 \times 48 - 2 \times 6 = 84$ in (a), and similarly 68 in (b) and 88 in (c). The resulting 'inflated' curves are 8-connected. Their covered areas are simply connected and have size $2 \times 152$ minus twice the overlapped segment length. \qed

\subsection{Boundary segment lengths}

This subsection contains an analysis of the lengths of the common segments of a minimal boundary ghost when overlapped with adjacent shifted ghosts. Let, as before, $v_1, v_2, \dots, v_n$ denote the start direction vectors for the tiles $U_{n}^a$.

In Section 5 of \cite{CekoGhosts} the lengths of the common subsegments have been computed for $V_n^a$. They are equal to the lengths of the common horizontal edges of $U_{n-1}^a$ and its adjacent copy. 
Denote the three segment lengths of $V_n^a$ by $s_n^1, s_n^2, s_n^3$. It follows from Theorem 5 of \cite{CekoGhosts} that they satisfy the following recursions.

\begin{equation} \label{vals}
s_{n+1}^1 = s_n^2 + 2s_n^3,~~s_{n+1}^2 = s_n^1,~~s_{n+1}^3=s_n^2,
\end{equation}
\vskip.1cm
\noindent with initial conditions $s_3^1 = 2$,~ $s_3^2=1$,~$s_3^3 = 0$. 

When boundary ghosts $V_n$ are overlapped using the vectors in Eq. \eqref{eq:Vinflatedbyd}, two of the three shift directions are different to those given by Eq. \eqref{eq:adjac}. Then the same ghost perimeter is cut into segments $s_n^{i'}$, with the same segment being $s_n^{2} =  s_n^{1'}$. The remainder of the same perimeter is cut into different lengths, as $s_n^{2'}$ and $s_n^{3'}$. Computed values for the recursive lengths of segments $s_n^{i'}$ are given in Eq. \eqref{eq:sprime_segmentlengths}.

\begin{equation} \label{eq:sprime_segmentlengths}
s_{n}^{1'} = s_{n-2}^{2'} + 2s_{n-3}^{3'},~~s_{n}^{2'} = s_n^{1'} + 4s_{n-2}^{3'},~~s_{n}^{3'} = s_{n-2}^{1'},
\end{equation}
\vskip.1cm

The segments $s_n^i$ and $s_n^{i'}$ of the perimeters of $V_n$ obtained by either method follow Eq.\eqref{eq:s_nformula} for $n \geq 5$, agree with the perimeter result of Eq. \eqref{eq:xnformula},

\begin{equation} \label{eq:s_nformula}
        s_{n}^i = s_{n-2}^i + 2s_{n-3}^i.
\end{equation}

Segment lengths, perimeters and areas for ghosts $V_n^a$ for 
$n=3-8$ are given in Table \ref{tab:Un_sn_pn_An}.

\begin{table}[h] \label{tab:VnSegmentsAreas}
\begin{minipage}{280pt}
\centering
\begin{tabular}{c@{ } c@{ } c@{ } c@{ } c@{ } c@{ } c@{ }c@{ } c@{ } c@{ } }
\toprule
$V_n^a~$&$(p_n,q_n)$~&$s_n^1$~&$s_n^2$~&$s_n^3$&$s_n^{1'}$~&$s_n^{2'}$~&$s_n^{3'}$~&$P_n$~~&$A_n$\\
\midrule
$V_3^a$~&(1,1)~&2&1&0&1&1&1&6&7 \\
$V_4^a$~&(-1,1)~&1&2&1&2&2&0&8&12  \\
$V_5^a$~&(-3,-1)~&4&1&2&1&5&1&14&23  \\
$V_6^a$~&(-1,-3)~&5&4&1&4&4&2&20&42  \\
$V_7^a$~&(5,-1)~&6&5&4&5&9&1&30&79  \\
$V_8^a$~&(7,5)~&13&6&5&6&14&4&48&152  \\
\botrule
\end{tabular}
\end{minipage}
\caption{Computed boundary segment lengths for ghosts made using $V_n^a$ with 3 to 8 zero-sum projections (the last vector of $V_n^a$ is $(p_n, q_n)$). The values for the three segment lengths, $s_n^i$, are followed by the segment lengths $s_n^{i'}$, the perimeter of the boundary ghost, $P_n$ and, finally, the covered area $A_n$. Here the values for segment lengths $s_n^i$ are given by \eqref{eq:adjac} and for $s_n^{i'}$ by \eqref{eq:Vinflatedbyd}, $P_n = 2\sum{s_n^i} = 2\sum{s_n^{i'}}$, $A_n = 2^{n-1}+P_n/2$.}\label{tab:Un_sn_pn_An}
\end{table}

\noindent {\bf Example 6.2.} For the tiling where the adjacent tiles of $U_7$ are obtained according to Eq.\eqref{eq:adjac}, the shift vectors are $\pm(-3,7)$, $\pm (17,3)$ and $\pm (14,10)$. Table \ref{tab:Un_sn_pn_An} shows the ghost $V_8^a$ is composed of six segments of length $13, 6, 5, 13, 6, 5$ with perimeter $P_8= 48$ and area $A_8 = 152$.

 The tiling of $V_8$ by the vectors of Eq.\eqref{eq:Vinflatedbyd} uses shift vectors $\pm(14,10)$, $\pm (10,-2)$ and $\pm (-4,-12)$, giving $V_8^a$ segment lengths $6, 14, 4, 6, 14, 4$ with the same perimeter $P_8= 48$ and area $A_8 = 152$. 
\qed

Using the shifts from Eq. \eqref{eq:adjac} for $V_8^b$, the perimeter halves are broken into the same segment lengths $13,6,5$ and $6, 14, 4$ for Eq. \eqref{eq:Vinflatedbyd}. For $V_8^{a'}$ and $V_8^{b'}$, the perimeter halves are broken into segment lengths $9,5,12$ and $12,2,12$ respectively.

The individual curved segments $s_n^i$ and $s_n^{i'}$ of a boundary ghost $V_n$ are not by themselves ghosts for $n$ projection directions. On their own, the segments generally have zero sums only for the  starting vectors $(0,1), (1, 0)$ for $V_n^b$ and $(0,1), (1,1)$ for $V_n^a$.

\subsection{Inflation by multiple ghosts} 
 \label{subsec:Vn_multiples}     

\begin{figure}
    \centering
    \includegraphics[width=0.4\linewidth]{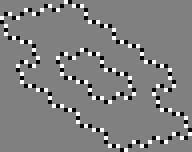}
    \includegraphics[width=0.35\linewidth]{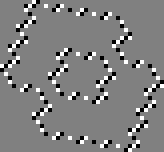}
    \caption{\hskip3cm(a) \hskip4cm(b)\\Images of the tiling using all six shifts of the boundary ghosts (a) $V_8$ shown as a reflected version  Fig. \ref{fig:ghosts_u71_v81}(b) and (b) $V_8^b$ shown in Fig. \ref{fig:ghosts_u72_v82}(b). The original ghosts (shown superimposed in the centre) are fully erased by the tiling of its six surrounding boundary ghosts.}
    \label{fig:ghostplus6copies}
\end{figure}

By multiple inflation we mean tiling several shifted boundaries at once. Suppose the original tile, $W(1)$, is added at $m-1$ other tile shifts to become a new, larger tile $W(m)$. Then the line sums of $W(m)$ become 0 in the same directions as was the case in the original boundary ghost $W(1)$.
The number of pixels of $W(m)$ equals $m$ times the number of pixels of $W(1)$ minus the total number $N$ of overlapped and hence cancelled boundary pixels. The number $N$ can be computed using the data for segment lengths, $s_n^i$ or $s_n^{i'}$. We give some examples of inflation by multiple ghosts.
\vskip.1cm
\noindent {\bf Example 6.3.} In Fig. \ref{fig:ghostplus6copies}(a) the inside boundary ghost is a reflected copy of the original tile $W_8(1) = V_8^a$, (with perimeter $48$ pixels and area $152$ pixels), as in Fig. \ref{fig:ghosts_u71_v81}(b). All six adjacent tiles are then added to it. The result is shown as the outside boundary tile $W_8(7)$.

Since in this case every adjacent tile contributes half of the perimeter of $W_8(1)$, the perimeter of $W_8(7)$ is exactly three times the perimeter of $W_8(1)$, that is $144 ~(= 3 \times 48)$. The area of $W_8(7)$ is somewhat less than $7 \times 152 = 1084$ pixels, more precisely, $7 \times 2^7 + 144:2 = 968$.

Fig. \ref{fig:ghostplus6copies}(b) shows the corresponding result where we start with $V_8^b$ from Fig. \ref{fig:ghosts_u72_v82}(b). The result is again a linear transformation of Fig. \ref{fig:ghostplus6copies}(a). \qed

\vskip.1cm

\noindent {\bf Example 6.4.} A minimal boundary ghost $W_n(1) = V_n$ with perimeter $P_n(1)$ can be inflated to yield boundary ghosts $W_n(m)$ of perimeter $3P_n(1)$ and covered area $m \times 2^n + 3P_n(1)/2$ in the following cases:

    $m=7$, the original tile is surrounded by all six adjacent tiles,
    
    $m=6$, any five among the six adjacent tiles are added to the original tile,
    
    $m=5$, any four among the six adjacent tiles are added to the original tile where one tile has both neighbours missing,
    
    $m=4$, any three among the six adjacent tiles are added to the original tile where each adjacent tile has both neighbours missing.

\begin{figure}
    \centering
    \includegraphics[width=0.8\linewidth]{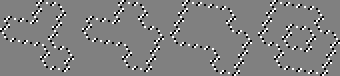}
    \caption{The ghost $V_8^b$ (with $48$ boundary pixels that enclose area 152 from \ref{fig:ghosts_u72_v82}(b)) can be inflated by adding three, four, five or six copies of itself. Each inflated ghost has the same perimeter of $3\times48 = 144$ boundary pixels (shown with 3 copies on the left, to 6 copies on the right). The ghost areas are, respectively, $584, 712, 840, 968$ pixels, being $m\times 2^7 + 144/2$). A copy of the original ghost $V_8^b$, for comparison, has been re-inserted at the centre of the ghost with six added copies, as shown on the right.}
    \label{fig:V8_4to7_tiles}
\end{figure}

There are a total of fifteen different variants of the above cases where each adjacent tile and each missing tile contributes half of the perimeter of the original tile to the perimeter of the inflated boundary ghost. Examples for $m = 3,4,5,6$ are shown in Fig. \ref{fig:V8_4to7_tiles}.

\begin{figure}
    \centering
    \includegraphics[width=0.3\linewidth]{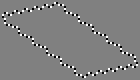}
    \includegraphics[width=0.425\linewidth]{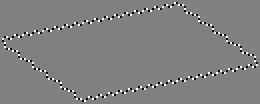}
    \caption{\hskip3cm(a) \hskip3cm(b)\\(a) Inflated boundary ghost, $W_8(9)$, constructed as nine copies of $W_8(1)$ by adding all adjacent tiles and two copies of $V_8$ at replacements $\pm (6,14)$. It consists of 160 pixels and covers area 1232. 
    (b) The inflated ghost image, $W_8(18)$, formed by adding image (a) to a copy of itself shifted by vector $(24,-8)$. It is composed of 220 pixels and covers area 2414.}
    \label{fig:ghostw81_w81pluscopy}
\end{figure}

The addition of extra shifts of an already inflated boundary ghost on to its own perimeter may be used, for example, to produce a better fitting ghost.

\vskip.1cm
\noindent {\bf Example 6.5} Fig. \ref{fig:ghostw81_w81pluscopy}(a) shows the inflated ghost after adding the additional ghost copies of the original $V_8$ at the vector displacements $\pm ((10,2) +(14,-10)) =\pm (24, -8)$. Note that again the internal boundary vanishes exactly when the boundary ghosts are overlapped in the right way. This inflated boundary ghost $W_8(9)$ has a perimeter of 144 from $W_8(7)$ augmented by $2 \times (4 + 4)$ equals 160, and therefore has area $9 \times 2^7 + 160/2 = 1232$. 

Fig. \ref{fig:ghostw81_w81pluscopy}(b) shows the inflated ghost image $W_8(18)$ formed by adding image (a) to a copy of itself shifted by vector $3 \times (10,2) = (30, 6)$. Its perimeter is $2 \times (160 - 3 \times 14 - 2 \times 4) = 220$ giving it area $18 \times 2^7 + 220/2 = 2414$. Both images in Fig. \ref{fig:ghostw81_w81pluscopy} retain the same 8 zero-sum projection directions and have a nearly parallelogram-shape. \qed

\section{Watermarking by boundary ghosts} \label{sec:watermarking}

The inflated boundary ghosts $W_n(m)$ may serve as useful watermarks when superimposed onto general forms of digital data.

A relevant example application might be to individually and uniquely stamp $N$ copies of a shared document or image on behalf of $N$ authorised owners. The aim is to prevent any unauthorised changes to the original data (by any of the $N$ owners or outsiders) and to prevent unauthorised ownership. We propose using $N$ differently inflated ghosts $W_n(m)$ as proprietary watermarks.

Assume the data is 8-bit and non-binary. The addition of $\pm1$ values should be checked for wrapped data values, to avoid the watermarked image leaving any detectable traces. $W_n$ should be spread almost edge to edge of the image being watermarked and be shaped and positioned to avoid image areas where the data is deemed to be visually more significant or sensitive.

As all ghost images have zero total sums, adding them to the test data does not change the sum of the test image. The sum of \emph{any} line projection for the angles of set $S_N$ will also not change.
  
Digital data can be discretely projected by, for example, the Mojette Transform (MT) \cite{guedonMojette} or the Finite Radon Transform (FRT) \cite{MatusFRT}. The FRT is restricted to image sizes $p \times p$, for $p$ prime. The projected data can be exactly reconstructed using projections made at any set of sufficient angles $S$. By the Katz criterion \cite{Katz}, a 2D projection set $S$ of $N$ angles is sufficient to exactly reconstruct an image of size $(n_x, n_y)$ pixels \emph{iff}, for angle set $S_N = (p_i, q_i), ~i = 1$ to $N$:

\begin{equation} \label{eq:Katz12}
    \sum_{i = 1}^{N} \mid{p_i}\mid \ge n_x , ~~\sum_{i = 1}^{N} \mid{q_i}\mid  \ge n_y. 
\end{equation}

\begin{figure}
    \centering
    \includegraphics[width=0.9\linewidth]{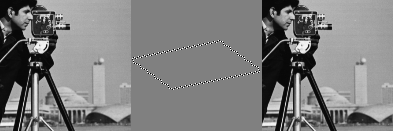}
   \caption{\hskip1cm (a) \hskip3cm (b) \hskip3cm (c) \\Watermarking example. (a) A $131 \times 131$ portion of an 8-bit integer test image, ‘cameraman’.(b) The inflated boundary ghost image $W_8(18)$ from Fig. \ref{fig:ghostw81_w81pluscopy}(b), white = +1, black = -1, else zero. (c) The watermarked data: image (b) added to test image (a).}
   \label{fig:cameraman_watermark}
\end{figure}

We present two ways of watermarking any digital image.                    
\vskip.1cm
\noindent {\bf Example 7.1} Fig. \ref{fig:cameraman_watermark}(a) is an example portion of an 8-bit grey level digital test image (cameraman), with size 131 x 131 pixels. A proposed watermark is the inflated ghost $W_8(18)$ as shown in Fig. \ref{fig:cameraman_watermark}(b). The watermarked test image is shown in Fig. \ref{fig:cameraman_watermark}(c). A minimally sufficient set of projection angles, $S_{41}$, was chosen to satisfy Eq.\eqref{eq:Katz12}, having absolute sums of 131 and 130, respectively. The angle set, $S_{41}$, by design, includes the set $V_8^a$
$$V_8 = [0, 1; 1, 0; 1, 1; -1, 1; 3, 1; 1, 3; -5, 1; 7, 5].$$ located in $S_{41}$ at indices [1, 2, 3, 4, 9, 11, 26, 34].

The MT projections at the angle set $W_8$ for the test image plus the ghost image additions are exactly the same projections as for the test image alone. However, the added ghost \emph{will} perturb the content of the projected data for all of the other ($41-8=33$) angles in $S_{41}$. The amount of perturbation depends on the $(x, y)$ location where one, or possibly several, copies of this watermark are added. For the sufficient angle set $S_{41}$, the MT projections have lengths ranging from 131 to 1561 pixels for $131 \times 131$ data. 

\begin{figure}
    \centering
    \includegraphics[width=0.5\linewidth]{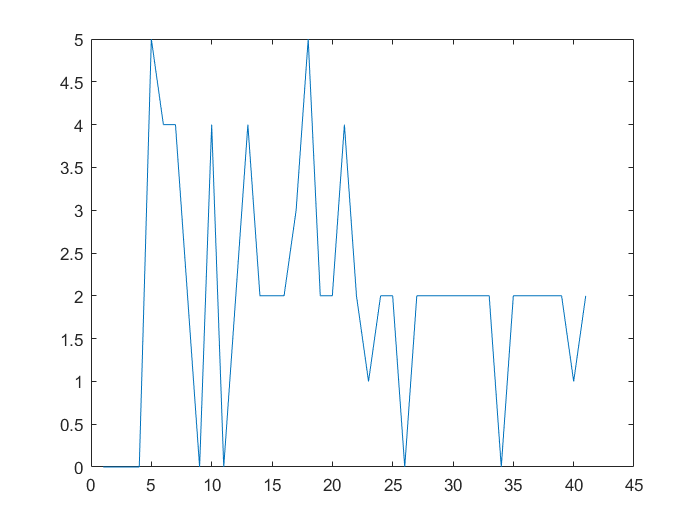}
    \caption{Maximum absolute difference values between the MT projected watermarked and original image data are zero for the $W_8$ boundary ghost angles, but are non-zero for all other angles. A secure (unaltered) copy of the watermarked test image will reproduce this pattern of differences.}
    \label{fig:MTproj_differences}
\end{figure}

Figure \ref{fig:MTproj_differences} plots the maximum absolute value of the differences between the original MT projections of Fig. \ref{fig:cameraman_watermark}(a) and the watermarked MT projections of image Fig. \ref{fig:cameraman_watermark}(c) for each angle in $S_{41}$. The zero differences are found only at the projection angles $[1, 2, 3, 4, 9, 11, 26, 34]$, which are the indices of $V_8^a$ within the set $S_{41}$. \qed

\begin{figure}
    \centering
    \includegraphics[width=\linewidth]{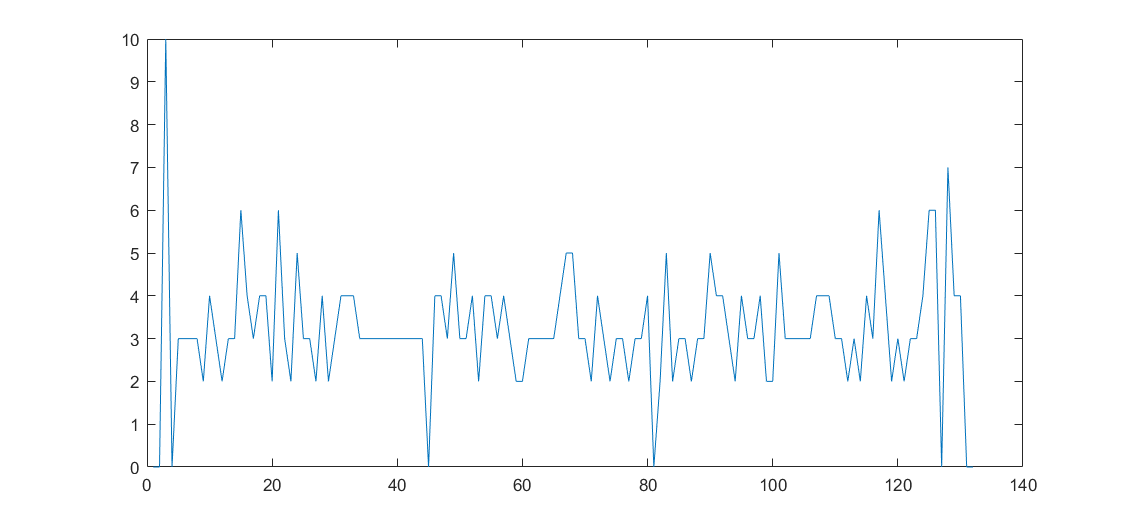}
    \caption{Maximum absolute difference values in the FRT projected watermarked and original image data are zero for the $W_8$ boundary ghost angles, but are non-zero for all other angles in the set $S_{132}$. A secure (unaltered) copy of the watermarked test image will reproduce this pattern of differences.}
    \label{fig:FRTproj_differences}
\end{figure}

\vskip.2cm
\noindent {\bf Example 7.2} The same watermarking method as in Example 7.1 can be applied using the FRT which, for $131 \times 131$ data, by default, has $132$ projection angles and each projection has the same fixed length, $131$. The FRT projection set, for any prime $p$, is automatically minimally sufficient to exactly reconstruct any $p \times p$ digital image. The angle set for $W_8$ occurs ‘naturally’ within the FRT angle set $S_{132}$, here at the FRT angle indices $[1, 2 , 4, 45, 81, 127, 131, 132]$. Those FRT projections of the test data also remain unchanged by the addition of the inflated ghost of Fig. \ref{fig:cameraman_watermark}(b), as shown in Fig. \ref{fig:FRTproj_differences}. \qed
\vskip.1cm
The non-ghost projection changes in Fig. \ref{fig:FRTproj_differences} are different (usually with larger values) than those of Fig. \ref{fig:MTproj_differences}. Discrete projections in the FRT sample the $p \times p$ image values in a cyclic (periodic) fashion, wrapping around the borders, to project the same image values into a fixed length $p$, the same length for all  $p+1$ angles.

To verify the authenticity of the data, the header of the image file watermarked by ghost $W_n$ can identify the set of angles used for the ghost by specifying which type of ghost ($V_n^a$, $V_n^{a'}$, $V_n^b$ or $V_n^{b'}$) was used. It should also record in the file header the maximum and/or minimum of the absolute values for the $W_n$ angle projections for the original data, as well as the $n$ maximum and/or minimum of the absolute projection values for either the $n$ angles of the ninety degree rotated ghost, or for the reflected ghost.

If the watermarked test data has been edited or otherwise tampered, the ghost projections would be altered from their authorised values, as would the non-ghost projections. If an extra copy of any valid inflated boundary ghost with the same set of projection angles (a ‘faked’ $W_n$ watermark) was later added to the already watermarked data, then the ghost projections would still not change, however the values for the non-ghost projections would no longer agree with the recorded values.

The resilience of $W_n$ as a watermark depends on how easily it can be removed or replaced from the host image data. Some form of local averaging may be tried to erase the $\pm1$ entries of $W_n$, but this is most unlikely to preserve the easily compared values for the projected images. 

Using method 1 in Subsection \ref{gendes} that inflates ghosts $U_{n-1}$ before applying the boundary shift to get $W_n$, has the advantage that a full-sized image of the tiling of $U_{n-1}$ can be overlaid on the image to be watermarked to plan a deliberate set of inflated steps that avoids making changes to sensitive areas of the image data.

\section{Boundary ghost shape variation}\label{sec:flexibility}

\begin{figure}
    \centering
    \includegraphics[width=0.9\linewidth]{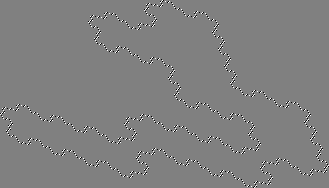}
    \caption{A boundary ghost, $W_{12}$, of size $329 \times 188$, with zero projections for 12 discrete directions. The ghost was built by a 'self-avoiding' random addition of the minimal boundary ghost $V_{12}^b$ shifted in one of its six nearest-neighbour inflation directions. The ghost has a perimeter of $1652$ pixels that enclose an area of $21,306$ pixels.}
    \label{fig:spiral_ghost_12 directions}
\end{figure}

The diversity of distinct boundary ghosts, such as $V_n^{a}, V_n^{a'}, V_n^{b}, V_n^{b'}$, may prove useful in secure watermark or digital signature applications. The construction of a binary boundary ghost has the following free variables:

1.	The start tiling and the start tile pattern $U_0$. The start tile was a single pixel in most of the examples shown here. However Ex. 2.2 provides an example of an arbitrarily selected start tile. The starting tile can be of size $n \times m $ with any tileable pattern of $\pm1$ pixel values that are simply connected.

2.	The boundary direction can be either $(\pm1, 1)$, $(1,0)$ and $0,1)$. Pixels differing by a multiple of the boundary direction should have the same pixel value.

3.	The choice for start vectors $v_1$ and $v_2$ such that shifts of $U_0$ by $v_1$ and by $v_2$ are adjacent tiles of $U_0$.

4.  The choice of $n$, the number of zero projection directions for the boundary ghosts (we show example ghosts here with $n$ from $7$ to $12$).

5.  At each of the first $n-1$ steps the choice between recurrence \eqref{eq:Vn1} and \eqref{eq:Vn2} (the $n^{th}$ step comes from adding the boundary direction). 

6.  Inflation. Arbitrary recursive addition of a copy of any boundary ghost $V_n$ to one, or more, of its six adjacent perimeter segments. Inflation can also be achieved using the minimal binary ghost $U_{n-1}$, as described in Subsection \ref{gendes}.

Further diversity can be achieved by inflating a mixture of $V_n^{a}, V_n^{a'}, V_n^{b}, V_n^{b'}$ ghosts. We have observed that segments of the boundaries of $a,a',b, b'$ ghosts (that are many pixels long) can be overlapped so as to cancel. The general set of vectors needed to tile such mixtures would be the subject of future work.

\vskip.1cm
\noindent {\bf Example 8.1.} Fig. \ref{fig:spiral_ghost_12 directions} shows a $329 \times 188$ image of a watermark that has zero projections in 12 directions. Here multiple copies of $V_{12}^b$ were added in one of the six neighbour directions, following a 'random-walk' pattern to create a larger ghost with a more complex shape. The number of unique watermark variations that are able to be made in this way grows exponentially.\qed

\vskip.1cm
\noindent {\bf Example 8.2.} Other binary ghosts, such as the minimum number of pixels ghost example $S_{10}$ (shown in Fig. \ref{fig:n10ghost40points}), may also be inflated by tiling. For such ghosts in $n>6$ directions, there are usually multiple distinct angle sets that have the same number of boundary points. For $n=7$, there are around a dozen different shaped binary ghosts (enclosing a wide range of different areas) that are each made from $20$ signed pixels. As a potential watermark, $S_{10}$ has the advantage of starting with the smallest possible number of boundary pixels for an $10$-directional ghost. Its strongly diagonal symmetry means the inflation directions are more restricted. To inflate $S_{10}$, the useful vector shifts are linear combinations of $(5,0)$ and $(0,5)$, such as $(10,15)$.  \qed

\section{Conclusion}\label{sec:Conclusion}

This work constructs binary images that vanish under discrete projection for $n$ specified directions. In particular, the presentation is focused on control of the size and shape of these 'ghost' images when they are enlarged (or 'inflated') by tiling the $2D$ plane with ghost copies, whilst always keeping zero projections for the same fixed set of $n$ directions. A potential application of these ghosts is for them to be embedded in digital data as 'indelible' and 'invisible' signatures or watermarks.

A new method is introduced to construct simply connected minimal binary ghosts. It generalises the earlier related work of \cite{CekoGhosts} by using an arbitrary signed tile ($T$) of pixels as a starting point, rather than beginning with just a single pixel.

For the case of a single starting pixel, we presented alternative sets of shift vectors that give new forms of minimal area ghosts $U_n^{a'}$ and new boundary ghosts, $U_n^{b}$ and $U_n^{b'}$. The larger choice of minimal boundary ghosts $V_n$ greatly increases the variety of uniquely different inflated ghosts $W_n$ that have the same $n$ projection directions.

We further investigated the segmented structure of these ghosts with a view to control the cancellation of segments of the ghost boundary as they are tiled by neighbouring copies. This cancellation process was used to keep small the count of perimeter pixels of the inflated ghosts, whilst enlarging the area encompassed by the tiled ghosts. The objective here was to create a large variety of 'inflated' boundary ghosts with uniquely distinct shapes composed from similar numbers of binary valued pixels.

The capacity to mould the shape of boundary projection ghosts may prove useful in the reconstruction of images from a small number of projection angles. In some forms of tomography, the acquisition of projections is limited to objects confined to a curved domain. In these cases, it will be possible to find configurations where, for certain (discrete) projection directions, boundary ghosts can be constructed that lie inside that domain. Then those projection directions do not help to resolve a reconstructed image.

The boundary ghosts considered here were assigned values of $\pm1$ to reduce their perceptibility in 8-bit integer image data. It is important to remember that the projections of binary ghosts remain invisible in their $n$ directions for either arbitrarily large or small values of any constant $\pm v$. For better imperceptibility in colour or hyper-spectral images, ghosts can be embedded across chromatic rather than in luminance channels (for example, by using $HSV$ or $LAB$ colour models, rather than adding directly into $RGB$ data). To watermark tables of float data, the value used for $v$ can be near epsilon, the required precision for float numbers under addition.

\section*{Declarations}

\begin{itemize}

\item Authors' contributions. The authors have no competing interests to declare that are relevant to the content of this article.

\item Acknowledgements. Andrew Tirkel, Scientific Technology, Brighton, Victoria, Australia, provided the authors with feedback on aspects relevant to data watermarking.

\item Availability of data and materials. All data pertaining to the numeric examples can be generated by the explicit constructions described within the manuscript.  
\end{itemize}

\bibliography{sn-bibliography}


\end{document}